\documentclass[amsmath,amssymb,prl,aps,mathbbm,superscriptaddress,twocolumn]{revtex4}
\usepackage{amssymb}
\usepackage{graphicx}
\usepackage{times}
\usepackage{subfigure}
\usepackage{float}

\begin{document}
    \title{Tripartite Genuine Non-Gaussian Entanglement in Three-Mode Spontaneous Parametric Down-Conversion}
    \author{A. Agust{\'\i}}
    \affiliation{Instituto de F{\'i}sica Fundamental, CSIC, Serrano, 113-bis, 28006 Madrid, Spain}
    \author{C.W. Sandbo Chang}
    \affiliation{Institute for Quantum Computing and Electrical and Computer Engineering, University of Waterloo, Waterloo, Ontario N2L 3G1, Canada}
    \author{F. Quijandr\'{\i}a}
    \affiliation{Microtechnology and Nanoscience, MC2, Chalmers University of Technology, SE-412 96 G\"oteborg, Sweden} 
    \author{G. Johansson}
    \affiliation{Microtechnology and Nanoscience, MC2, Chalmers University of Technology, SE-412 96 G\"oteborg, Sweden}
    \author{C.M. Wilson}
    \affiliation{Institute for Quantum Computing and Electrical and Computer Engineering, University of Waterloo, Waterloo, Ontario N2L 3G1, Canada}
    \author{C. Sab{\'\i}n}
    \affiliation{Instituto de F{\'i}sica Fundamental, CSIC, Serrano, 113-bis, 28006 Madrid, Spain}

    \email{soyandres2@gmail.com}
    \begin{abstract}
        We show that the states generated by a three-mode spontaneous parametric down-conversion (SPDC) interaction Hamiltonian possess tripartite entanglement of a different nature to other paradigmatic three-mode entangled states generated by the combination of two-mode SPDC interactions. While two-mode SPDC generates Gaussian states whose entanglement can be characterized by standard criteria based on two-mode quantum correlations, these criteria fail to capture the entanglement generated by three-mode SPDC. We use criteria built from three-mode correlation functions to show that the class of states recently generated in a superconducting-circuit implementation of three-mode SPDC ideally have tripartite entanglement, contrary to recent claims in the literature. These criteria are suitable for triple SPDC but we show that they fail to detect tripartite entanglement in other states which are known to possess it, which illustrates the existence of two fundamentally different notions of tripartite entanglement in three-mode continuous-variable systems.
    \end{abstract}

    \maketitle

    Parametric amplification of the quantum vacuum in superconducting-circuit architectures \cite{nation} has proven to be a very fruitful paradigm for quantum technologies. For instance, the high-frequency modulation of a superconducting quantum interference device (SQUID) terminating a superconducting transmission line can generate pairs of photons out of the vacuum \textemdash a particular realization of the dynamical Casimir effect \cite{dce} \textemdash which exhibit entanglement and other forms of quantum correlations \cite{entanglementchalmers, discord, steering, coherence}. These correlations become resources that can be used in many applications of quantum technologies, for instance, entangling distant qubits \cite{felicetti, andres} in distributed quantum computing architectures~\cite{qinternet}. 
    
    However, the use of these resources is limited by their bipartite nature, meaning the correlations only span two systems. Extending the entanglement to more modes would unlock access to a large number of new protocols including boson sampling \cite{bsamp}, the generation of microwave cluster states \cite{cluster}, quantum state sharing \cite{tristateshare}, quantum secret sharing \cite{qsecret1, qsecret2}, and quantum teleportation networks \cite{qtele}. One strategy to accomplish this is multitone modulation of the SQUID, with frequencies addressing multiple pairs of modes.
Theory predicts that this approach can produce genuine multipartite entanglement \cite{sorinnat,sorinpra} and it has recently been experimentally validated for three modes \cite{praappl}. While the demonstrated entanglement was genuinely tripartite, it was generated by the simultaneous action of a pair of two-mode interactions and was detected purely through the measurement of second-order correlations~\cite{tehreid}. It stands to reason that a single three-mode Hamiltonian might be better suited for the task of generating tripartite entanglement.

    In fact, a three-mode spontaneous parametric down-conversion (SPDC) Hamiltonian can be engineered in superconducting circuits by suitably flux pumping an asymmetric SQUID terminating a coplanar waveguide resonator, as recently demonstrated experimentally in \cite{inprogress}. As this scheme includes a direct three-mode interaction, the relevant physical features cannot be captured by second-order correlations, making it necessary to include higher-order correlations in the characterization of the state \cite{inprogress}. The presence of independent higher-order correlations is often referred to as non-Gaussianity in the system. Common second-order criteria, including those previously mentioned~\cite{tehreid}, fail to detect multipartite entanglement in these states, as was noted in \cite{threephotnoent}. This has led to the impression that three-mode SPDC may not be a useful quantum resource. Nevertheless, in this Letter,  we show that three-mode SPDC does produce entanglement, as well as the necessity to use higher-order correlations to detect the generated tripartite entanglement. Therefore, the claim in \cite{threephotnoent} that there is no entanglement in these states is overly broad. As we will prove below, the correct statement is that entanglement is, indeed, generated, but that it is non-Gaussian in nature.

    In this work, we use entanglement criteria based on third- and fourth-order correlations to detect tripartite entanglement in the class of states produced experimentally by three-mode SPDC~\cite{inprogress}. We show that the class exhibits both full inseparability and genuine tripartite entanglement. We also show that the same criteria fail to detect tripartite entanglement in states produced by quadratic Hamiltonians, in which the multimode interaction is induced by the combination of two-mode interactions. Since we know the latter also include states with genuine tripartite entanglement, as shown experimentally in \cite{praappl}, our results clearly suggest that higher-order SPDC interactions generate a different kind of multipartite entanglement, distinguished from the Gaussian entanglement most commonly studied in continuous-variable systems. We will refer to this novel notion of entanglement as genuine non-Gaussian entanglement. 

    Let us now start with the description of our results. We analyze a system related to the experimental setup of \cite{inprogress}, consisting of a superconducting resonator terminated by an asymmetric SQUID. We consider three field modes with frequencies $\omega_i$, ($i=a,b,c$) and the corresponding creation and annihilation operators $i,\,i^{\dagger}$ with standard bosonic commutation relations. We assume that initially each mode is in a weak thermal state $\rho_i (n_{th}^i)$, characterized by the corresponding low average number of thermal photons according to its frequency and temperature, as given by $\langle n_{th}^i\rangle=1/(e^{\beta\omega_i}-1)$, where $\beta\omega_i=\hbar\omega_i/(k_B T) \gg 1$.

    The system evolves under the interaction Hamiltonian:
    \begin{eqnarray}
        H_{I}&=& \hbar g_0\cos{\omega_0 t} (e^{i\theta_a}a+e^{-i\theta_a}a^{\dagger}) (e^{i\theta_b}b+e^{-i\theta_b}b^{\dagger})\nonumber\\ & &(e^{i\theta_c}c+e^{-i\theta_c}c^{\dagger}),
        \label{norwa-H}
    \end{eqnarray} 
    where $\theta_i$ are locally controllable phases and $g_0$ is the coupling strength. Choosing the coupling modulation $\omega_0$ as
    \begin{equation}
        \omega_0=\omega_a+\omega_b+\omega_c
    \end{equation}
    gives rise to the effective Hamiltonian, with a derivation resembling a rotating-wave approximation (RWA), that has the form in the interaction picture of:
    \begin{equation}
        H_{I}= \frac{\hbar g_0}{2} (e^{i\theta}abc+e^{-i\theta}a^{\dagger}b^{\dagger} c^{\dagger}),
        \label{rwa-H}
    \end{equation}
    where $\theta=\theta_a+\theta_b+\theta_c$, and in the following we fix it to zero, as it plays no interesting role in entanglement generation. This is the three-mode SPDC Hamiltonian \cite{sam,samcarl,hillary} required.

    The standard criteria to detect tripartite entanglement, such as \cite{tehreid,vanlockfuro}, are based on inequalities concerning expectation values and correlations which involve of course the three modes but in a pairwise fashion, such as $\langle x_i x_j\rangle$ ($x_i$, $x_j$ being the position quadratures associated to the modes, e.g., $x_i=(i + i^{\dagger})/\sqrt{2} $). However, looking at the Hamiltonian (\ref{rwa-H}) it seems natural to think that these criteria are not suitable in this case. Indeed it was shown in \cite{threephotnoent} that some of these criteria were not able to detect tripartite entanglement for these states. Perhaps the most compelling evidence proving this point is that the covariance matrix of an initial thermal state \textemdash including the vacuum\textemdash{} evolved under Hamiltonian (\ref{rwa-H}) remains diagonal. That is, Hamiltonian (3) does not produce any second-order correlations. But we must stress that the criteria used in \cite{threephotnoent} are sufficient but not necessary conditions on entanglement, and as such they are inconclusive when they fail. Thus, what is needed is to look for higher-order criteria able to capture the pure three-mode nature of the states generated by the Hamiltonian (\ref{rwa-H}). This nature has been demonstrated, both theoretically and experimentally, by the absence of second-order correlations together with the existence of third-order ones \cite{inprogress}.

    A typical approach to tripartite entanglement is to study correlations between all the possible bipartitions of the system. We recall that the definition of an entangled system is a system whose density matrix, $\rho$, is neither separable nor a mixture of separable states, that is $\rho\neq\sum_{i}P_i\rho_i^{(1)}\otimes\rho_i^{(2)}$, where each $\rho_i^{(1)}$ spans the first system and each $\rho_i^{(2)}$ the second. For instance, for each bipartition we can look at the inequalities developed in \cite{viethillery}. As usual, if the state is not entangled between the two subsystems, correlations between them are, by definition, classical and the inequalities hold. But if they are violated, we can conclude the state has to be entangled between those subsystems. If we define $A^{(1)}$ and $A^{(2)}$ as operators acting respectively on the Hilbert spaces of two subsystems in which the total system is split, by \cite{viethillery} if the total state is not entangled with respect to this partition, then:
    \begin{equation}
        |\langle A^{(1)} A^{(2)}\rangle| \leq \sqrt{\langle A^{(1)\dagger} A^{(1)}\rangle \langle A^{(2)\dagger} A^{(2)}\rangle}.\label{ineq} 
    \end{equation}
    Therefore, in our case, choosing the annihilation operator as the reference operator in all cases, we have that if condition
    \begin{equation}
        |\langle abc\rangle| \leq \sqrt{\langle N_i\rangle \langle N_j N_k\rangle},\label{ineq} 
    \end{equation}
    is violated for all three possible $i-jk$ bipartitions (namely $a-bc$, $b-ac$, $c-ab$) of the system then we know that the state is not biseparable with respect to any bipartition. In the above, $N$ is the number operator. If the state is not biseparable for the three bipartitions, then the state is said to be fully inseparable. Defining $I_i=|\langle abc\rangle| - \sqrt{\langle N_i\rangle \langle N_j N_k\rangle}$, we have that the state is fully inseparable if $I_i>0$ for the three bipartitions.

    However, even if the state has full inseparability, there is still the possibility that
    \begin{eqnarray}
        \rho &=&
        P_1\rho^{(a)}_{1}\otimes\rho^{(bc)}_{1}
        +
        P_2\rho^{(b)}_{2}\otimes\rho^{(ac)}_{2} \nonumber \\
        &+& 
        P_3\rho^{(c)}_{3}\otimes\rho^{(ab)}_{3}
        \label{bisep}
    \end{eqnarray}
or in other words, the state is a mixture of biseparable states (which implies $P_1 + P_2 + P_3 = 1$). In this particular state, it is as if the tripartite correlations were classical and, hence, we do not refer to full inseparability as a form of tripartite entanglement. Then it becomes immediate to define genuine tripartite entanglement as the correlations of fully inseparable states that cannot be written as (\ref{bisep}) \cite{tehreid,genuine1,genuine2}.  Note that we are in an analogous situation as before, looking for a condition every state like (\ref{bisep}) must follow, and concluding that any state violating it has to be genuinely entangled. Note too that the difference between full inseparability and genuine entanglement is only relevant for mixed states.

    We have not found in the literature a condition for genuine tripartite entanglement involving correlations of more than two modes. However, we can derive an inequality involving  $|\langle abc\rangle|$ that every state of the form (\ref{bisep}) follows. Using (\ref{bisep}) and the triangle inequality, it is straightforward to write
    \begin{eqnarray}
        |\langle abc\rangle_{\rho}|\leq P_1|\langle abc\rangle_{\rho_1}| + P_2 |\langle abc\rangle_{\rho_2}| +P_3|\langle abc\rangle_{\rho_3}|, \label{ineq1}
    \end{eqnarray}
    where in the lhs the expectation value refers to the total state $\rho$ while in the rhs refer to the different elements of the convex sum, that we are denoting $\rho_1$, $\rho_2$, $\rho_3$, namely
    \begin{eqnarray}
        \rho_1&=&\rho^{(a)}_{1}\otimes\rho^{(bc)}_{1}, \quad
        \rho_2 = \rho^{(b)}_{2}\otimes\rho^{(ac)}_{2} \nonumber \\
        \rho_3&=&\rho^{(c)}_{3}\otimes\rho^{(ab)}_{3}
    \end{eqnarray}
    \begin{figure}[H]
        \includegraphics[scale=0.6]{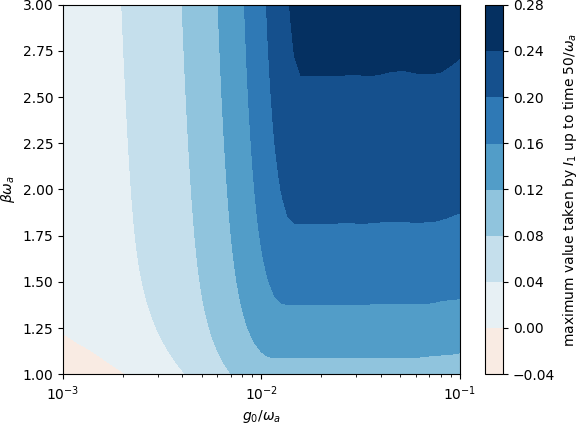}
        \caption{Maximum value of the witness $I_1$ taken in the time interval $(0,50\omega_a)$, when the system evolves under Hamiltonian (\ref{norwa-H}) and represented for several values of coupling $g_0$ and temperature $T = \hbar/k_B\beta$, in units normalized by the lower energy mode frequency, $\omega_a$. $I_2$ and $I_3$ have the same behavior, accounting for different mode frequencies. This indicates the system is fully inseparable for any value of $g_0$ and $\beta$ in the regime explored, except for very high temperature and low coupling.}
    \label{norwa-maxI1}
    \end{figure}

    Now we know that, by construction, $\rho_1$, $\rho_2$ and $\rho_3$ are biseparable and therefore they must follow the inequalities (\ref{ineq}). Therefore we have
    \begin{eqnarray}
        |\langle abc\rangle_{\rho}|&\leq& P_1\sqrt{\langle N_a\rangle_{\rho_1}\langle N_bN_c\rangle_{\rho_1}} \nonumber \\&+& P_2 \sqrt{\langle N_b\rangle_{\rho_2}\langle N_aN_c\rangle_{\rho_2}} \nonumber \\&+& P_3 \sqrt{\langle N_c\rangle_{\rho_3}\langle N_aN_b\rangle_{\rho_3}} . \label{ineq2}
    \end{eqnarray}
    Finally, using again (\ref{bisep}) we have that, for instance,
    \begin{eqnarray}
        P_1\langle N_a\rangle_{\rho_1}&=& \langle N_a\rangle_{\rho}- P_2\langle N_a\rangle_{\rho_2} - P_3\langle N_a\rangle_{\rho_3}\nonumber\\ &\leq &\langle N_a\rangle_{\rho}, \label{3forineq3}
    \end{eqnarray}
    and similarly with all the expectation values in (\ref{ineq2}). Putting everything together, we find that if the state is of the form (\ref{bisep}) then: 
    \begin{eqnarray}
        |\langle abc\rangle|&\leq& \sqrt{\langle N_a\rangle\langle N_bN_c\rangle} +  \sqrt{\langle N_b\rangle\langle N_aN_c\rangle} \nonumber \\&+& \sqrt{\langle N_c\rangle\langle N_aN_b\rangle}, \label{ineqfin}
    \end{eqnarray}
    where we have let the subindex $\rho$ drop since it would be the same for all expectation values. Therefore, we conclude that if a state  violates (\ref{ineqfin}), then it possesses genuine tripartite entanglement. If it does not violate the inequality (\ref{ineqfin}) but violates (\ref{ineq}) for the three bipartitions, then it is just fully inseparable. We define $G= |\langle abc\rangle|- \sqrt{\langle N_a\rangle\langle N_bN_c\rangle} - \sqrt{\langle N_b\rangle\langle N_aN_c\rangle} \nonumber -\sqrt{\langle N_c\rangle\langle N_aN_b\rangle}$, and then $G>0$ is the condition for genuine tripartite entanglement.

    With these witnesses $I_i$ and $G$, we can begin to study the entanglement generated by three-mode SPDC. If we consider the initial temperature negligible, then the initial state is the vacuum. If in addition to this we suppose that we are in a perturbative regime where the system evolves to a  pure state containing only the vacuum and a triplet with small probability amplitude $\alpha$, then $\langle abc\rangle\simeq \alpha$, while the $\langle N_i\rangle$, $\langle N_jN_k\rangle$ are of order $|\alpha|^2$. Therefore, for very low temperatures and coupling strengths, the conditions for entanglement are expected to be satisfied. 

    In order to confirm and generalize this analytical intuition,  we present now numerical results for the above inequalities for the states generated by the evolution under the interaction Hamiltonian (\ref{rwa-H}) for three modes $\omega_a=\omega$, $\omega_b=2\omega$, $\omega_c=3\omega$ in the parameter regime $\beta_i >1$, $g_0/\omega<<1$ (low temperature and low coupling).

\begin{figure}
    \includegraphics[scale=0.6]{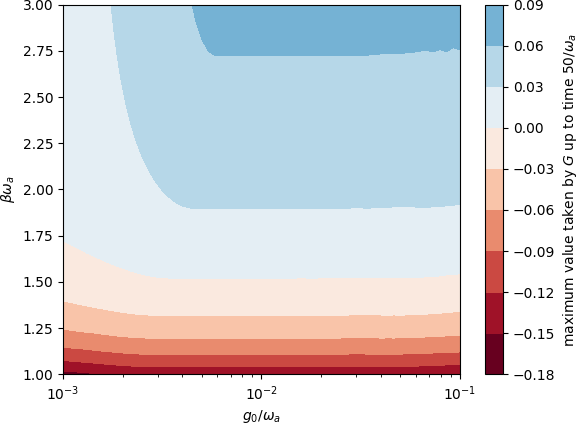}
    \caption{Maximum value of the witness $G$ taken in the same time interval as $I_1$ in Fig. (\ref{norwa-maxI1}) and represented against the same variables in the same units and conditions. As it can be seen, genuine nongaussian entanglement is reported for low temperatures over $\beta\omega_a = 1.6$ (blue/darker upper region) while the witness fails to capture the system's entanglement, if any, for higher temperatures comparable to level spacing (red/darker lower region).}
    \label{norwa-maxG}
\end{figure}

    We find that the system is fully inseparable in all the explored parameter regime, except for very low coupling and high temperature ($g_0 < 0.002\omega_a$ and $\beta\omega_a < 1.2$). This can be seen in Fig. (\ref{norwa-maxI1}), where we present the maximum value of $I_1$ over the time interval $(0,50/\omega_a)$ as a function of coupling and temperature. This proves $a-bc$ inseparability. The values of $I_2$ and $I_3$ are not shown, but, in fact, they generally exceed $I_1$. We expect this behavior since the single modes in their bipartitions have higher frequencies which leads to fewer thermal photons and a better parametric amplification of vacuum by the Hamiltonian. Genuine entanglement is detected for temperatures lower than $\beta\omega_a = 1.6$, as reported in Fig. (\ref{norwa-maxG}). Summarizing, almost everywhere in the explored parameter regime (that is, $g_0 < 0.1\omega_a$ and $\beta<1$) the system contains tripartite correlations, but only for temperatures below $\beta\omega_a = 1.6$ are those correlations known to be genuine entanglement.

    \begin{figure}[H]
        \includegraphics[scale=0.6]{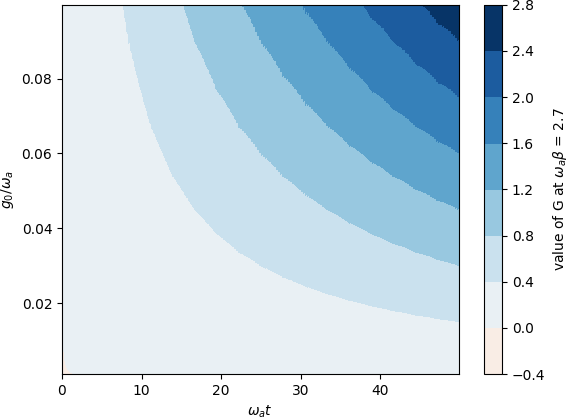}
        \caption{Value of the witness $G$ as a function of time and coupling, in units of the lowest frequency mode when the temperature is $k_BT = \hbar\omega_a/2.7$, well in the genuinely entangled regime shown in Fig. (\ref{norwa-maxG}) and close to vacuum. The system evolves under the RWA Hamiltonian (\ref{rwa-H}). As expected, the value of the witness increases both with time and coupling.}
        \label{rwa-Gvsg0}
    \end{figure}

    The role of the coupling is richer than the temperature, as higher coupling can require higher-order corrections in perturbation theory, as well as break the rotating-wave approximation, leading to a discrepancy between the full Hamiltonian in Eq. (\ref{norwa-H}) and the effective one in Eq. (\ref{rwa-H}). Intuitively, the coupling controls the rate of evolution of the interacting system.  Figure (\ref{rwa-Gvsg0}) shows how the system evolves from a slightly thermal state under the RWA Hamiltonian (\ref{rwa-H}), developing an ever increasing value of $G$ with a rate determined by the coupling. 
  
\begin{figure}[H]
        \includegraphics[scale=0.6]{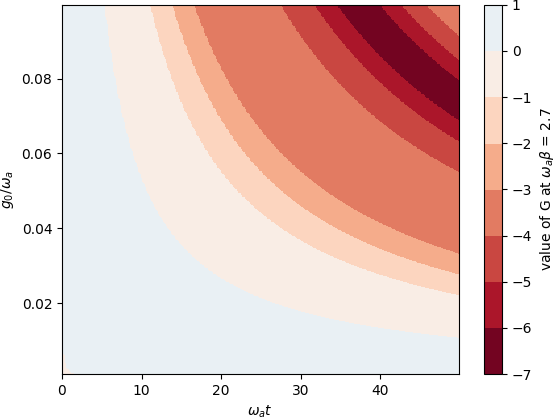}
        \caption{Value of the witness $G$ under the same conditions as in Fig. (\ref{rwa-Gvsg0}) but evolving under the full Hamiltonian (\ref{norwa-H}). In the perturbative regime, i.e. for short times or low coupling, the prediction for $G$ is the same as in Fig. (\ref{rwa-Gvsg0}). However, outside this regime the behavior is significantly different with the value of the witness becoming negative, for longer times and higher couplings.}
        \label{norwa-Gvsg0}
    \end{figure}

  However, we may expect that the Hamiltonian (\ref{rwa-H}) will stop being an effective description of the full Hamiltonian (\ref{norwa-H}) at high couplings and long times. That is, the RWA may break down.  In fact, we observe this breakdown as shown in Fig. (\ref{norwa-Gvsg0}), which plots the same information but evolving the system under the full Hamiltonian (\ref{norwa-H}). In this new scenario, the behavior of $G$ is the same as Figure (\ref{rwa-Gvsg0}) for short times (in units of the coupling), indicating genuine entanglement. However, $G$ becomes negative after some time. This behavior is expected, since the RWA neglects terms that do not create or annihilate photon triplets, e.g., $a^\dagger b c$, $a^\dagger b^\dagger c$. Therefore, when their effects become relevant, we expect weaker third-order correlations. In the experimental setup of \cite{inprogress}, the reported correlations follow from the evolution of the system under their RWA Hamiltonian, therefore the relevant parameter regime in this work has to be contained in the region with $G>0$ in Figure (\ref{norwa-Gvsg0}). In addition to this, the timescale during which the RWA is valid is short compared to dissipative ones, so we conclude that decoherence in the system will not produce a relevant change in $G$ before the full Hamiltonian (\ref{norwa-H}) spoils it.

    We have seen, then, that the states generated by the action of a three-mode SPDC Hamiltonian such as (\ref{norwa-H}) or (\ref{rwa-H}) evolving from an initial weakly thermal state possess tripartite entanglement \textemdash contrary to the claim in \cite{threephotnoent}\textemdash which can be detected by our three-mode criteria. Now we compare these results with the case of double two-mode SPDC (that is, a Hamiltonian of the form $ab + ac + a^\dagger b^\dagger +a^\dagger c^\dagger$). In \cite{praappl}, it was shown that the resulting state possesses not only full inseparability, but also genuine entanglement by means of entanglement tests based on second-order correlations. We report that $G$ fails at detecting any entanglement at any coupling or temperature in the regime in which it would be expected to (that is, $\beta > 1$ and $g_0/\omega_a < 0.1$). Therefore, we are now in the opposite scenario: tripartite entanglement is detected by two-mode criteria but not by our three-mode criteria. This suggests that $G$ and the conditions explored in \cite{threephotnoent, praappl} detect two different kinds of entanglement. Hence, we label the entanglement signaled by $G > 0$ as tripartite genuine \textit{non-Gaussian} entanglement, given the clear non-Gaussian nature of states evolved under either Eqs. (\ref{norwa-H}) or (\ref{rwa-H}).

    Summarizing, we have shown that three-mode SPDC interaction Hamiltonians generate states with both full inseparability and genuine tripartite entanglement when acting upon a weak thermal state, contrary to previous claims in the literature. The type of tripartite entanglement displayed by these states is different from other paradigmatic three-mode states, and therefore needs to be captured by different entanglement criteria. We introduce  entanglement criteria based on three-mode correlations and show that our states satisfy them in a promising parameter regime. However, we show that double-SPDC Hamiltonians acting on weak thermal states, which generate states that have been proven to also possess tripartite entanglement by means of different second-order criteria, fail to satisfy our conditions. This points to two different classes of continuous-variable tripartite entanglement in three-mode systems.

    Our results pave the way for multipartite entanglement tests in the experimental setup of \cite{inprogress} and could be a guide for the characterization and measurement of entanglement in three-mode SPDC in other platforms \cite{tospdc,tospdc2}.

    A.A. and C.S. have received financial support through the Postdoctoral Junior Leader Fellowship Programme from la Caixa Banking Foundation (LCF/BQ/LR18/11640005). C.M.W. and C.W.S.C. acknowledge funding from NSERC of Canada and the Canada First Research Excellence Fund (CFREF). F.Q. and G.J. acknowledge support from the Wallenberg Center for Quantum Technology (WACQT)
    
\end{document}